# Putting Lipstick on Pig: Enabling Database-style Workflow Provenance

Yael Amsterdamer[2], Susan B. Davidson[1], Daniel Deutch[3], Tova Milo[2], Julia Stoyanovich[1], Val Tannen[1]

[1]University of Pennsylvania, USA
{susan, jstoy, val}@cis.upenn.edu

[2]Tel Aviv University, Israel
{yaelamst, milo}@cs.tau.ac.il

[3]Ben Gurion University, Israel
deutchd@cs.bgu.ac.il

## ABSTRACT

Workflow provenance typically assumes that each module is a "black-box", so that each output depends on all inputs (*coarse-grained* dependencies). Furthermore, it does not model the *internal state* of a module, which can change between repeated executions. In practice, however, an output may depend on only a small subset of the inputs (*fine-grained* dependencies) as well as on the internal state of the module. We present a novel provenance framework that marries database-style and workflow-style provenance, by using Pig Latin to expose the functionality of modules, thus capturing internal state and fine-grained dependencies. A critical ingredient in our solution is the use of a novel form of *provenance graph* that models module invocations and yields a compact representation of fine-grained workflow provenance. It also enables a number of novel graph transformation operations, allowing to choose the desired level of granularity in provenance querying (ZoomIn and ZoomOut), and supporting "what-if" workflow analytic queries. We implemented our approach in the *Lipstick* system and developed a benchmark in support of a systematic performance evaluation. Our results demonstrate the feasibility of tracking and querying fine-grained workflow provenance.

## 1. INTRODUCTION

Data-intensive application domains such as science and electronic commerce are increasingly using workflow systems to design and manage the analysis of large datasets and to track the provenance of intermediate and final data products. Provenance is extremely important for verifiability and repeatability of results, as well as for debugging and trouble-shooting workflows [10, 11].

The standard assumption for workflow provenance is that each module is a "black-box", so that each output of the module depends on all its inputs (*coarse-grained* dependencies). This model is problematic since it cannot account for common situations in which an output item depends only on a small subset of the inputs (*fine-grained* dependencies). For example, the module function may be *mapped* over an input list, so that the $i^{th}$ element of the output list depends only on the $i^{th}$ element of the input list (see Taverna [18, 29]). Furthermore, the model does not capture the internal state of a module, which may be modified by inputs seen in previous executions of the workflow (e.g., a learning algorithm), and an output may depend on some (but not all) of these previous inputs. Maintaining an "output depends on all inputs" assumption quickly leads to a very coarse approximation of the actual data dependencies that exist in an execution of the workflow; furthermore, it does not show the way in which these dependencies arise.

For example, consider the car dealership workflow shown in Figure 1. The execution starts with a buyer providing her identifier and the car model of interest to a bid *request* module that distributes the request to several car *dealer* modules. Each dealer looks in its database for how many cars of the requested model are available, how many sales of that model have recently been made, and whether the buyer previously made a request for this model, and, based on this information, generates a bid and records it in its database state. Bids are directed to an *aggregator* module that calculates the best (minimum) bid. The user then makes a *choice* to accept or decline the bid; if the bid is accepted, the relevant dealership is notified to finalize the purchase. If the user declines the bid but requests the same car model in a subsequent execution, each dealer will consult its bid history and will generate a bid of the same or lower amount.

Coarse-grained provenance for this workflow would show the information that was given by the user to the bid request module, the bids that were produced by each dealer and given as input to the aggregator, the choice that the user made, and which dealer made a sale (if any). However, it would not show the dependence of the bid on the cars that were available at the time of the request, on relevant sale history, and on previous bids. Thus, queries such as "Was the sale of this VW Jetta affected by the presence of a Honda Civic in the dealership's lot?","Which cars affected the computation of this winning bid?", and "Had this Toyota Prius not been present, would its dealer still have made a sale?" would not be supported. Coarse-grained provenance would also not give detailed information about how the best bid was calculated (a minimum aggregate).

Finer-grained provenance has been well-studied in database research. In particular, a framework based on *semiring annotations* has been proposed [17], in which every tuple of the database is annotated with an element of a provenance semiring, and annotations are propagated through query evaluation. For example, semiring addition corresponds to alternative derivation of a tuple, thus, the union of two relations corresponds to adding up the annotations of tuples appearing in both relations. Similarly, multiplication corresponds to joint derivation, thus, a tuple appearing in the result of a join will be annotated with the product of the annotations of the two joined tuples. The provenance annotation captures the way in which the result tuple has been derived from input tuples. Note that the present paper focuses on *data manipulation* and not on module boundaries or execution order. The recorded provenance therefore allows only limited queries about module invocations and flow, and only when these have a direct effect on the data. For instance, a workflow execution for an empty bid request will not appear in the provenance. *The overall contribution of this paper is a framework that marries database-style and workflow provenance models, capturing internal state as well as fine-grained dependencies in workflow provenance.*

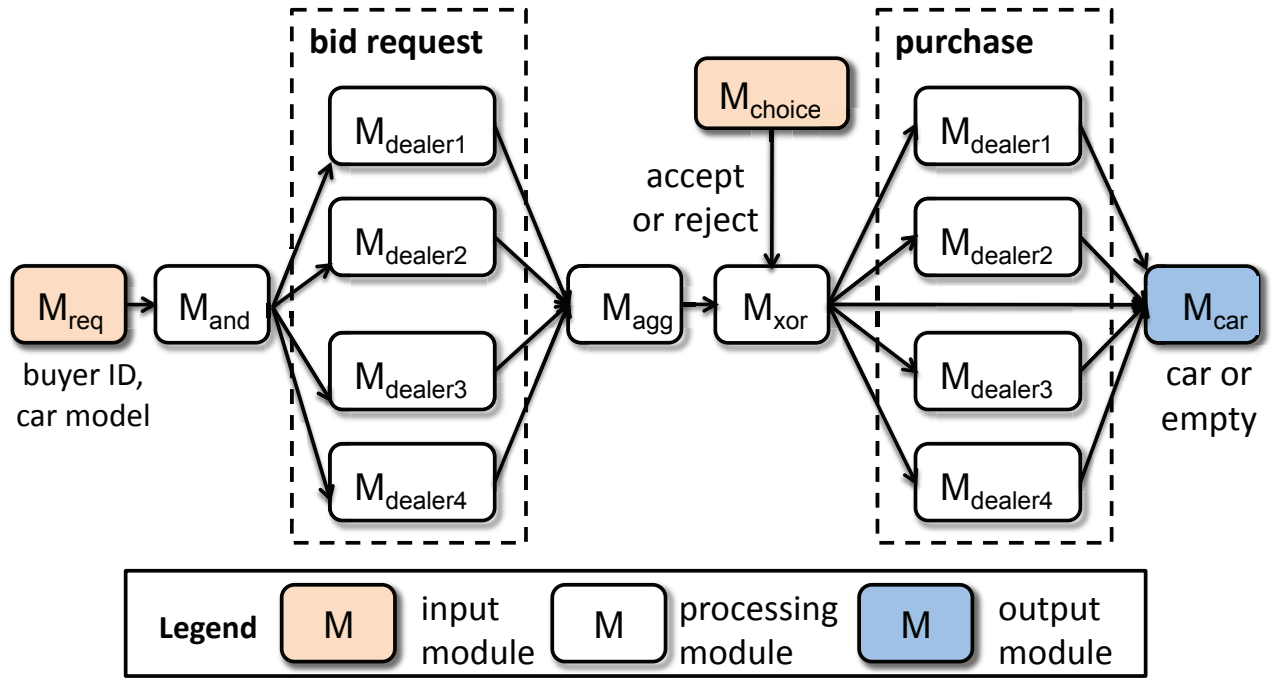

**Figure 1: Car dealership workflow.**

The framework uses Pig Latin to expose the functionality of workflow modules, from which provenance expressions can be derived. Pig Latin is increasingly being used for analyzing extremely large data sets since it has been "designed to fit in a sweet spot between the declarative style of SQL, and the low-level, procedural style of map-reduce" [26]. Pig Latin's use of complex, nested relations is a good match for the data types found throughout data-oriented workflows, as is the use of aggregates within expressions.

Note that it may not be possible to completely expose the functionality of a module using Pig Latin. Returning to our example, the bid generated by a dealer is calculated using a complex function that can only be captured in Pig Latin with a User Defined Function (UDF). In this case, coarse-grained provenance must be assumed for the UDF portion of the dealer expression. In contrast, fine-grained provenance for the functionality of the aggregator module can be exposed using aggregation. The framework therefore allows module designers to expose collection-oriented data processing, while still allowing opaque complex functions.

Several challenges arise in developing this framework. First, we must develop a notion of fine-grained provenance for individual modules that are characterized by Pig Latin expressions. We can do this by translating Pig Latin expressions into expressions in the bag semantics version of the nested relational calculus (NRC) [7] augmented with aggregation. Thus, we derive provenance from the framework of [2, 14]. *The development of a provenance framework for Pig Latin expressions is the first specific contribution of this paper.*

Second, fine-grained provenance information for a workflow may become prohibitively large if maintained in a naive way since a workflow may contain tens of modules, and may have been executed hundreds of times. A critical ingredient in our solution is the ability to reduce the potentially overwhelming amount of fine-grained provenance information using a novel form of *provenance graph*. While this idea was used in [16] for positive relational algebra queries, our provenance graph representation also accounts for aggregation, nested relational expressions, and module invocations, resulting in a much richer provenance graph model. *The second contribution of the paper is the development of a comprehensive and compact graph-based representation of fine-grained provenance for workflows, which also captures module invocations and module state changes.*

Third, since fine-grained workflow provenance yields a much richer graph model than the standard used for workflows (the Open Provenance Model [23]) or what is used for databases in [16], a richer set of queries can be asked. We thus define the graph transformation operations ZoomIn, ZoomOut and deletion propagation, and show how they can be used to answer novel workflow analysis queries. For example, we demonstrate how users can go between fine- and coarse-grained views of provenance in different portions of the workflow using ZoomIn and ZoomOut, and how deletion propagation may be used to answer "what-if" queries, e.g., "What would have been the bid by dealer 1 in response to a particular request if car $C_2$ were not present in the dealer's lot?". These graph transformations can be used in conjunction with a provenance query language such as ProQL [20]. *The third contribution of the paper is the definition of graph transformation operations ZoomIn, ZoomOut and deletion propagation, which enable novel workflow analysis queries.*

Finally, having presented a data model and query primitives for fine-grained workflow provenance, we develop the *Lipstick* system that implements provenance tracking for Pig Latin and supports provenance queries. We also propose a performance benchmark that enables systematic evaluation of *Lipstick* on workflows with a variety of topologies and module implementations. We show, by means of an extensive experimental evaluation, that tracking and querying fine-grained provenance is feasible. *The fourth and final contribution of this paper is the development of the* Lipstick *system and of an experimental benchmark.*

**Related Work.** Workflow provenance has been extensively studied and implemented in the context of systems such as Taverna [18], Kepler [5], Chimera [13], Karma [28], and others. These systems keep coarse-grained representation of the provenance, and many conform to OPM [23]. Ideas for making workflow provenance information more fine-grained have recently started to appear. Some examples include [29] which gives a semantics for Taverna 2 that allows specifying how input collection data are combined (e.g., "dot" or "cross" product), [22] that considers the representation and querying of this finer-grained provenance, and COMAD-Kepler [5] that considers provenance for collection-oriented workflows. In all of these works, however, data dependencies are explicitly declared rather than automatically generated from the module functionality specification. Moreover, unlike the present work, these works do not include a record of *how* the data is manipulated by the different modules (for instance, aggregation), nor do they capture module inner state. The same holds for Ibis [25], where different granularity levels can be considered for data and process components, but the link between data and process components captures only which process components generated which data items, with no record of the computational process that lead to the result, i.e., a simple form of "why"-provenance [8] is captured. PASSv2 [24] takes a different and very general approach, which combines automatic collection of system-level provenance with making an API available to system developers, who can then code different provenance collection strategies for different layers of abstraction.

The workflow model used in this paper is inspired by work on modeling data centric Web applications [12] (which does not deal with provenance). The use of nested relations and of Pig Latin, rather than of the relational model, allows a natural modeling for our target applications. We use a simpler control flow model than does [12]; extending our results to a richer flow model is left for future research.

Data provenance has also been extensively studied for query languages for relational databases and XML (see, e.g., [3, 6, 9, 14, 17]); specifically, in this paper we make use of recent work on provenance for aggregate queries [2]. Our modeling of provenance as a graph is based on [20]. The line of work that is based on semirings, starting from [17], was proven to be highly effective, in the context of data provenance, for applications such as deletion propagation, trust assessment, security, and view maintenance. Consequently, we believe that using this framework as a foundation for fine-grained workflow provenance will allow to support similar applications in this context.

Several recent works have attempted to marry workflow provenance and data provenance. In [1] the authors present a model based on *provenance traces* for NRC; in [19] the authors study provenance for map-reduce workflows. We also mention in this context the work of [21] that shows how to map provenance for NRC queries to the Open Provenance Model (although it does not consider workflow provenance per-se; their input is simply an NRC query). However, these models lack the structuring and granularity levels of our model, and naturally lack the corresponding query constructs introduced here. Another advantage of our approach is that it is based on the foundations given in [2, 14, 17], opening the way to the applications described above.

**Paper Outline.** In Section 2 we give an overview of Pig Latin and the semiring provenance model of [14, 15, 17] and describe our workflow model. In Section 3 we show how to generate provenance graphs for Pig Latin expressions and for full workflow executions. Section 4 presents our provenance query language that uses fine-grained provenance for answering complex analysis tasks. Section 5 describes the implementation of the *Lipstick* prototype and of our proposed performance evaluation benchmark, and presents results of an experimental evaluation, demonstrating the practicality of our approach. We conclude in Section 6.

# 2. PRELIMINARIES

We start with a brief overview of Pig Latin, then define our model of workflows and their executions, and conclude with an overview of the semiring framework for data provenance.

## 2.1 Pig Latin Primer

Pig Latin is an emerging language that combines high-level declarative querying with low-level procedural programming and parallelization in the style of map-reduce. (Pig Latin expressions are compiled to map-reduce.) We review some basic features of the language, see [26] for details.

*Data.* A Pig Latin *relation* is an unordered bag of tuples. Relations may be *nested*, i.e., a tuple may itself contain a relation. A Pig Latin relation is similar to a standard nested relation, except that it may be heterogenous, i.e., its tuples may have different types. For simplicity we will only consider homogenous relations in this paper, but our discussion can be extended to the heterogenous case.

*Query constructs.* We now review a fragment of the language that will be used in the sequel.

- **Arithmetic operations.** Pig Latin supports standard arithmetic operations such as `SUM`, `MAX`, `MIN`, etc. When applied to a relation with a single attribute, the semantics is that of *aggregation* (no grouping).
- **User Defined Functions (UDFs).** Pig Latin allows calls to (external) user defined functions that take relations as input and return relations as output.
- **Field reference (projection).** Fields in a Pig Latin relation may be accessed by position (e.g., `R.$2` returns the second attribute of relation `R`) or by name (e.g., `R.f1` returns the attribute named `f1` in `R`).
- `FILTER BY`. This is the equivalent of a select query; the semantics of the expression `B=FILTER A BY COND` is that `B` will include all tuples of `A` that correspond to the boolean condition `COND`.
- `GROUP`. This is the equivalent of SQL group by, without aggregation. The semantics of `B=GROUP A BY f` is that `B` is a nested relation, with one tuple for each group. The first field is `f` (unique values), and the second field `A` is a bag of all `A` tuples in the group.
- `FOREACH A GENERATE f1, f2,...,fn, OP(f0)` does both projection and aggregation. It projects out of `A` the attributes that are not among `f0,f1,...fn` and it `OP`-aggregates the tuples in the bag under `f0` (which is usually built by a previous `GROUP` operation).
- `UNION`, `JOIN`, `DISTINCT`, and `ORDER` have their usual meaning.

Pig Latin also includes constructs for *updates*. However we ignore these in the sequel, noting that the state-of-the-art for update provenance is still insufficiently developed.

*Relationship with (bag) NRC.* A key observation is that Pig Latin expressions (without UDFs) can be translated into the (bag semantics version of the) nested relational calculus (NRC) [7]. Details will be given in an extended version of this paper but we note here that this translation is the foundation for our provenance derivation for Pig Latin.

## 2.2 Model

We start by defining the notion of a module before turning to workflows and their execution.

The functionality of a module is described by Pig Latin queries. The queries map relational inputs to outputs but may also use and add to the module's relational *state*, which may affect its operation when the module is invoked again.

DEFINITION 2.1. *A* module *is identified by a unique name and is specified using a 5-tuple* $(S_{\text{in}}, S_{\text{state}}, S_{\text{out}}, Q_{\text{state}}, Q_{\text{out}})$, *where* $S_{\text{in}}, S_{\text{out}}$ *and* $S_{\text{state}}$ *are (disjoint) relational schemas, while* $Q_{\text{state}} : S_{\text{in}} \times S_{\text{state}} \to S_{\text{state}}$ *(state manipulation) and* $Q_{\text{out}} : S_{\text{in}} \times S_{\text{state}} \to S_{\text{out}}$ *are Pig Latin queries.*

EXAMPLE 2.1. *Our example in Figure 1, associates with each dealership a module $M_{\text{dealer}k}, k = 1, 2, \ldots$. These modules have the same specification, but different identities. Each of them receives different inputs, namely bid requests from potential buyers, which are instances of the following common input schema $S_{\text{in}}$:*

Requests

| UserId | BidId | Model |
|---|---|---|

*Each module $M_{\text{dealer}k}$ maintains a distinct state, which includes cars that are available and cars that were sold at the dealership. Each such state is an instance of the following state schema $S_{\text{state}}$ (several temporary relations omitted):*

Cars

| CarId | Model |
|---|---|

SoldCars

| CarId | BidId |
|---|---|

InventoryBids

| BidId | UserId | Model | Amount |
|---|---|---|---|

*The output schema $S_{\text{out}}$ of these modules is:*

Bids

| Model | Price |
|---|---|

*The modules $M_{\text{dealer}k}, k = 1, 2, \ldots$ share the same state manipulation and output query specification, but the queries act on different $S_{\text{in}}, S_{\text{state}}, S_{\text{out}}$-instances, corresponding to each module. Note that each $M_{\text{dealer}k}$ is invoked twice during workflow execution, first to place bids in response to requests and second to handle a purchase. We omit the code that switches between these two functionalities and the code for purchases, and show only the more interesting portion of the state manipulation query $Q_{\text{state}}$ that handles bid requests:*

```
ReqModel = FOREACH Requests GENERATE Model;
Inventory = JOIN Cars BY Model, ReqModel BY Model;
SoldInventory = JOIN Inventory BY CarId,
                     SoldCars BY CarId;
CarsByModel = GROUP Inventory BY Model;
SoldByModel = GROUP SoldInventory BY Model;
NumCarsByModel = FOREACH CarsByModel GENERATE
      group as Model, COUNT(Inventory) as NumAvail;
NumSoldByModel = FOREACH SoldByModel GENERATE
      group as Model, COUNT(SoldInventory) as NumSold;
AllInfoByModel = COGROUP Requests BY Model,
                    NumCarsByModel BY Model,
                    NumSoldByModel BY Model;
InventoryBids = FOREACH AllInfoByModel GENERATE
FLATTEN(CalcBid(Requests,NumCarsByModel,NumSoldByModel));
```

A Pig Latin join produces two columns for the join attribute, e.g., a join of `Cars` and `ReqModel` on `Model` creates columns `Cars::Model` and `ReqModel::Model` in `Inventory`, with the same value. We refer to this column as `Model`, and omit trivial projection and renaming queries from $Q_{state}$. The last statement in $Q_{state}$ invokes a user-defined function `CalcBid`, which, for each tuple in `AllInfoByModel`, returns a bag containing one output tuple; we use `FLATTEN` to remove nesting, i.e., to return a tuple rather than a bag.

Multiple modules may be combined in a workflow. A workflow is defined by a Directed Acyclic Graph (DAG) in which every node is annotated with a module identifier (name), and edges pass data between modules. The data should be consistent with the input and output schemas of the endpoints, and every module must receive all required input from its predecessors. An exception is a distinguished set of nodes called the *input nodes* that have no predecessors and get their input from external sources.

DEFINITION 2.2. *Given a set $\mathcal{M}$ of module names, a workflow is defined by $W = (V, E, L_V, L_E, In, Out)$, where*

- *$(V, E)$ is a connected DAG (directed acyclic graph).*
- *$L_V$ maps nodes in $V$ to module names in $\mathcal{M}$ (the same module may be used multiple times in the workflow).*
- *$L_E$ maps each edge $e = (v_1, v_2)$ to one or more relation names that belong to both $S_{\text{out}}$ of $L_V(v_1)$ and $S_{\text{in}}$ of $L_V(v_2)$. The relations names assigned to two adjacent incoming edges are pairwise disjoint.*
- *$In \subseteq V$ is a set of* input nodes *without incoming edges and $Out \subseteq V$ is set of* output nodes *without outgoing edges.*
- *Moreover, we assume that all module inputs receive data, i.e., for each node $v \in V - In$ the $S_{\text{in}}$ of $L_V(v)$ is included in $\bigcup_{e=(v',v)} L_E(e)$.*

The restriction to acyclicity is essential for our formal treatment. Dealing with recursive workflows would introduce potential non-termination in the semantics and, to the best of our knowledge, this is still an unexplored area from the perspective of provenance. This does *not* prevent modules from being executed multiple times, e.g., in a loop or parallel (forked) manner; however looping must be bounded. Workflows with bounded looping can be unfolded into acyclic ones, and are thus amenable to our treatment.

EXAMPLE 2.2. *In the car dealership workflow of Figure 1, input and output nodes are shaded, and the module name labeling a node, $L_v$, is written inside the node. The workflow start node corresponds to the input module $M_{\text{request}}$, through which potential buyers can submit their user ids and the car models of interest. This information, together with an indication that this is a bid request, is passed to four dealerships, denoted by $M_{\text{dealer1}}, \ldots, M_{\text{dealer4}}$, whose functionality was explained above. These modules each output a bid, and the bids are given as input to an aggregator module, $M_{\text{agg}}$, which calculates the best (minimum) bid. The user then accepts or declines the best bid. If the bid is accepted, the relevant dealership is notified ($M_{\text{xor}}$) so that it can update its state (*`SoldCars`*). The purchased car information or an empty result is output by the module $M_{\text{car}}$.*

Given concrete instances for the input relations of the input nodes we can define a workflow *execution.* With this we can define a *sequence of executions* corresponding to a sequence of input instances.

DEFINITION 2.3. *A single execution of a workflow $W = (V, E, L_V, L_E, In, Out)$, given a workflow input (instances for the input relations of modules in $L_V(In)$) and a workflow state (instances for state relations of each module in $L_V(V)$), is obtained by choosing a topological order $[v_0, ..., v_k]$ of the DAG $(V, E)$ and for each $i = 0, ..., k$, in order:*

- *Executing the state manipulation query and the output query of the module $L_V(v_i)$, on its input and current state instances and obtaining new state instances as well as output instances for the module.*
- *For each edge $e = (v_i, v_j) \in E$, and each relation name $R$ in $L_E(e)$, copying the instance of $R$ which is an output for $L_V(v_i)$ into the instance of $R$ which is an input for $L_V(v_j)$.*

*The output of this execution consists of the resulting instances for the output relations of the modules in $L_V(Out)$. Moreover, the execution also produces a new state for each module since each module invocation may change its state.*

*Given a sequence $\mathbf{I}_0, \ldots, \mathbf{I}_n$ of workflow inputs and an initial workflow state $\mathbf{S}_0$, a corresponding sequence of executions for $W$, $\mathbf{E}_0, \ldots, \mathbf{E}_n$, is such that for $i = 0, ..., n$ $\mathbf{E_i}$ is the execution of $W$ given $\mathbf{I}_i$ and $\mathbf{S}_i$, and producing the output $\mathbf{O}_i$ and the state $\mathbf{S}_{i+1}$. The overall sequence produces $\mathbf{O}_0, \ldots, \mathbf{O}_n$.*

Each choice of a topological ordering defines a *reference semantics* for the workflow. While implementations may use parallelism, we assume that acceptable parallel implementations must be *serializable* (cf. the general theory of transactions) and therefore their input-output semantics must be the same as one of the reference semantics defined here.

Note that our modeling of workflow state allows module invocations to affect the state used by subsequent invocations of the *same* module, within the same execution as well as subsequent executions.

We now show part of an execution of our sample workflow.

EXAMPLE 2.3. *Let us assume that at the beginning of the execution some cars already exist in the inventory, and that the state of the module $M_{\text{dealer1}}$ contains:*

Cars

| CarId | Model |
|---|---|
| $C_1$ | *Accord* |
| $C_2$ | *Civic* |
| $C_3$ | *Civic* |

*We also assume no cars were sold and no bids were made. Now, fix a topological order of the modules, e.g., $M_{\text{request}}, M_{\text{and}}, M_{\text{dealer1}}, M_{\text{dealer2}}, M_{\text{dealer3}}, ..., M_{\text{car}}$. Consider the application of $M_{\text{dealer1}}$ during that execution, with the input:*

Requests

| UserId | BidId | Model |
|---|---|---|
| $P_1$ | $B_1$ | *Civic* |

*Then the state update query $M_{\text{dealer1}}$ (described above) is executed. To track the stages of the query execution, we show the generated intermediate tables.*

ReqModel

| Model |
|---|
| *Civic* |

Inventory

| CarId | Model |
|---|---|
| $C_2$ | *Civic* |
| $C_3$ | *Civic* |

SoldInventory

| CarId | Model | BidId |
|---|---|---|

CarsByModel

| Model | Inventory |
|---|---|
| *Civic* | $\{\langle C_2, Civic\rangle, \langle C_3, Civic\rangle\}$ |

SoldByModel

| Model | SoldInventory |
|---|---|

NumCarsByModel

| Model | NumAvail |
|---|---|
| *Civic* | 2 |

NumSoldByModel

| Model | NumSold |
|---|---|

AllInfoByModel

| Model | Requests | NumCarsByModel | NumSoldByModel |
|---|---|---|---|
| *Civic* | $\{\langle P_1, B_1, Civic\rangle\}$ | $\{\langle Civic, 2\rangle\}$ | {} |

InventoryBids

| BidId | UserId | Model | Amount |
|---|---|---|---|
| $B_1$ | $P_1$ | *Civic* | \$20$K$ |

*The value of the bid is then the module output. If this bid is the minimal among all bids (as determined by $M_{\text{agg}}$), and if the user accepts this bid (via the input module $M_{\text{choice}}$), a car from the first dealership will be sold to the user. After the execution of $M_{\text{agg}}$, the* `SoldCars` *table will contain:*

SoldCars

| CarId | BidId |
|---|---|
| $C_2$ | $B_1$ |

*Otherwise, it will remain empty. Things also works well with a sequence of executions corresponding to a sequence of requested bids: after each execution, the state of each dealership module $M_{\text{dealer}k}$ has an up-to-date inventory which is part of the initial state of the next execution in the sequence.*

## 2.3 Data Provenance

In Section 3 we will develop a provenance formalism and show how provenance propagates through the operators of Pig Latin. This formalism is based on the semiring framework of [14, 15, 17] and on its extension to aggregation and group-by developed in [2], which we now briefly review.

Given a set $X$ of *provenance tokens* with which we annotate the tuples of input relations, consider the (commutative) semiring $(\mathbb{N}[X], +, \cdot, 0, 1)$ whose elements are multivariate *polynomials* with indeterminates (variables) from $X$, with coefficients from $\mathbb{N}$ (natural numbers), and where $+$ and $\cdot$ are the usual polynomial addition and multiplication. It was shown in [14, 17] that these polynomials capture the provenance of data propagating through the operators of the positive relational algebra and those of NRC (with just base type equality tests). Intuitively, the tokens in $X$ correspond to "atomic" provenance information, e.g., tuple identifiers, the $+$ operation corresponds to *alternative use* of data (such as in union and projection), the $\cdot$ operation corresponds to *joint use* of data (as in Cartesian product and join), 1 annotates data that is always available (we do not track its provenance), and 0 annotates absent data. All this is made precise in [17] (respectively [14]), where operators of the relational algebra (NRC) are given semantics on relations (nested relations) whose tuples are annotated with provenance polynomials. In this paper we use an alternative formalism based on graphs and therefore we omit the definitions of the operations on (nested) annotated relations.

In [2] we have observed that the semiring framework of [17] cannot adequately capture aggregate queries. To solve the problem we have further generalized $\mathbb{N}[X]$-relations by extending their data domain with *aggregated values*. For example, in the case of `SUM`-aggregation of a set of tuples, such a value is a *formal sum* $\sum_i t_i \otimes v_i$, where $v_i$ is the value of the aggregated attribute in the $i^{th}$ tuple, while $t_i$ is the provenance of that tuple. We can think of $\otimes$ as an operation that "pairs" values with provenance annotations. A precise algebraic treatment of aggregated values and the equivalence laws that govern them is based on semimodules and tensor products and is described in [2]. Importantly, in this extended framework, relations have provenance also as part of their values, rather than just in the tuple annotations.

Another complication is due to the semantics of group-by as it requires exactly one tuple for each occurring value of the grouping attribute — an implicit duplicate elimination operation. To preserve correct bag semantics, we annotate grouping result tuples with $\delta(t_1 + \cdots + t_n)$, where $t_1, \ldots, t_n$ are the provenances of the tuples in a group, and the unary operation $\delta$ captures duplicate elimination.

# 3. PROVENANCE FOR WORKFLOWS

We next present the construction of provenance graphs for workflow executions, which will be done in two steps. We start with a coarse-grained provenance model similar to the standard one for workflows [23], but enriched with some dedicated structures that will be useful in the sequel. Then, we extend this model to fine-grained provenance, detailing the inner-workings of the modules.

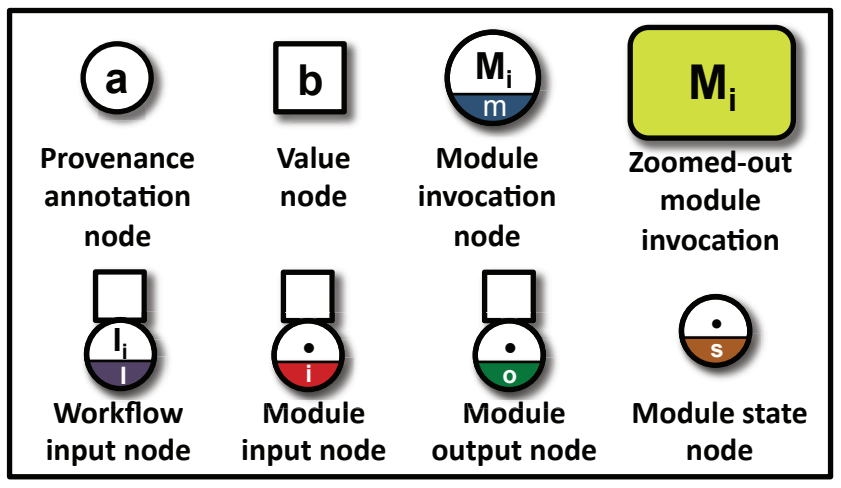

(a) Legend

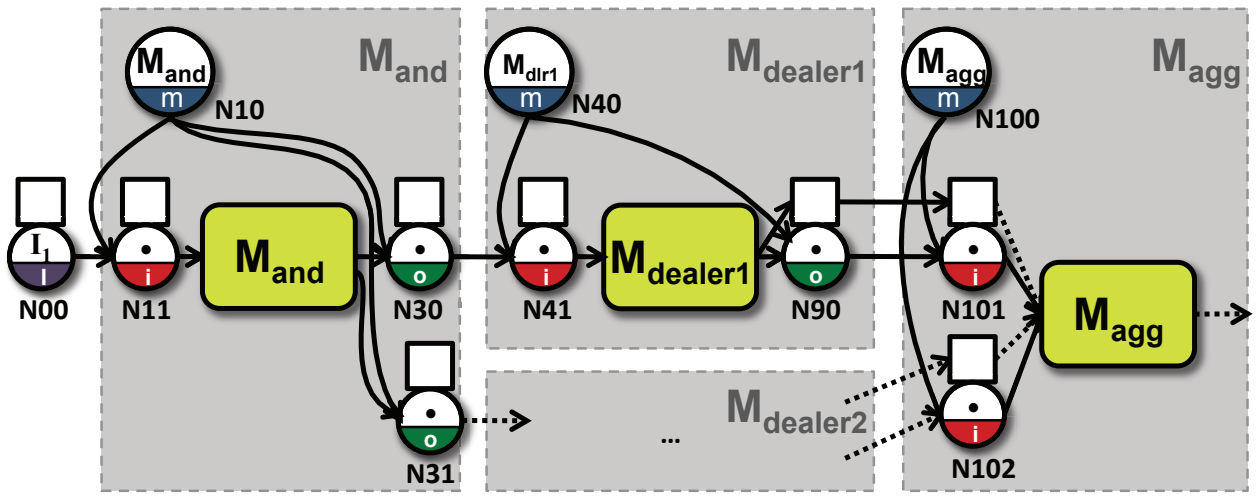

(b) Coarse-grained provenance

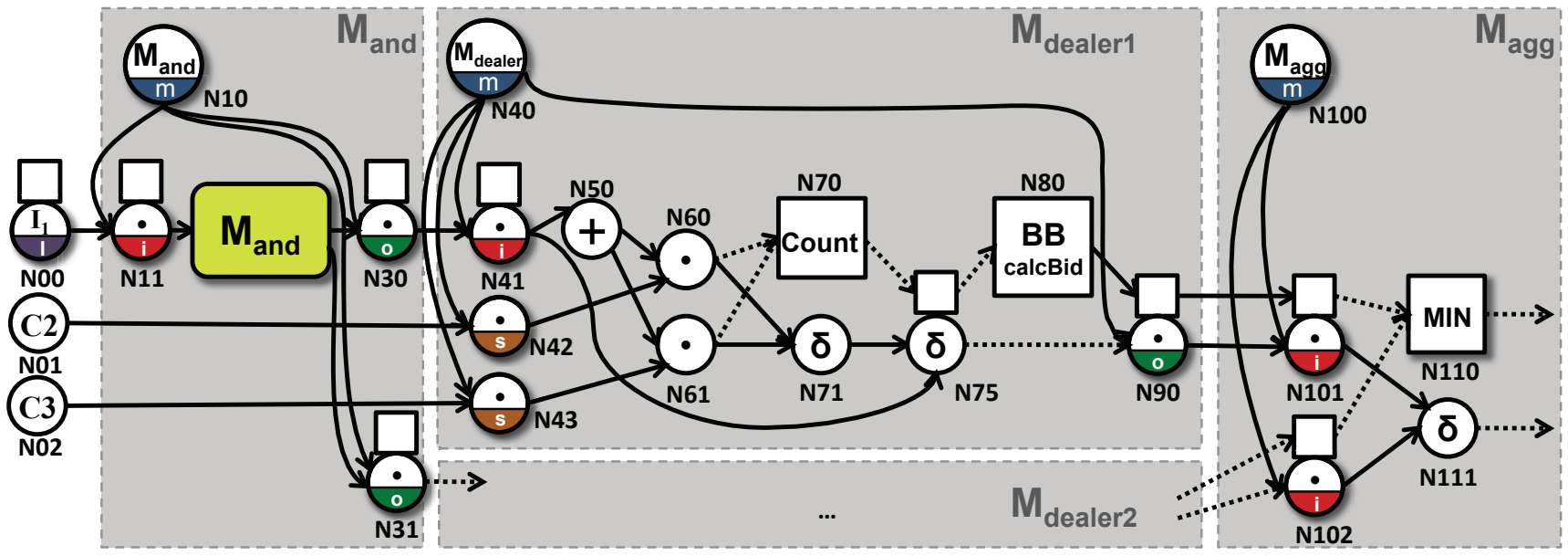

(c) Fine-grained provenance

**Figure 2: Partial provenance graphs for the car dealership workflow.**

In Section 4 we will formalize the connection between coarse and fine-grained provenance and describe querying provenance at flexible granularity levels.

## 3.1 Coarse-grained Provenance

Coarse-grained provenance describes the sequence of module invocations in a particular workflow execution (or a sequence of executions), their logical flow and their input-output relations. Figure 2(b) shows coarse-grained provenance for the car dealership (Figure 1); different kinds of nodes are given in the legend (Figure 2(a)). We only give details for $M_{\text{dealer1}}$, as all dealer modules are the same.

**Provenance and value nodes.** We distinguish between provenance nodes (p-nodes, represented by circular nodes in the figure), and nodes representing values (v-nodes, represented by square nodes). Both kinds of nodes must appear in the graph following the mixed use of values and provenance annotations for aggregate queries (see Section 3.2). To reduce visual overload, we will sometimes use a composite node (square on top of a circle) to denote both provenance and value nodes of the same tuple. See, e.g., $N_{41}$.

**Workflow Input nodes.** For each workflow input tuple, provided by some input module (e.g., $M_{\text{request}}$), we create a p-node of type "i" (for "input"). For example, $N_{00}$ represents the provenance of a bid request.

**Module invocation nodes.** For each invocation of a module $M$ we create a new $M$-labeled node of type "m". For example, $N_{40}$ represents the first invocation of $M_{\text{dealer1}}$.

**Module input nodes.** For each tuple given as input to some module $M$, we create a new p-node of type "i", labeled with the semiring $\cdot$ operation (see Section 3.2). We connect to this node the p-node of the tuple, as well as the module invocation p-node. The operation $\cdot$ is used here, in its standard meaning of joint derivation, to indicate that the flow relies jointly on both the actual tuple and on the module. Similarly, we create a v-node of type "i" for every value of the input tuple that appears in the graph. See, e.g., the node $N_{41}$, representing the input of $M_{\text{dealer1}}$.

**Module output nodes.** A construction similar to that of module input nodes, but with node type "o".

**Zoomed-out module invocation nodes.** For each module invocation, all input and output nodes are connected to a single node of this kind, shown by a rounded rectangle in Figure 2(b). These nodes are replaced by a detailed description of internal computations in fine-grained provenance, discussed next.

## 3.2 Fine-grained Provenance

Coarse-grained provenance gives the logical flow of modules and their input-output relations, but hides many other features of the execution, such as a module's state DBs, operations performed, and computational dependencies between data. We next consider fine-grained provenance that allows "zooming-into" modules to observe these features.

Our definition of fine-grained workflow provenance is based on the provenance polynomials framework for relational algebra queries, and its extension to handle aggregation, introduced in Section 2.3. However, we use graphs rather than polynomials to represent provenance. Provenance tokens and semiring operations such as $\cdot$, $+$, and $\delta$, are used as labels for nodes in the provenance graph. For example, an expression $t_1 + t_2$ is represented by three nodes labeled $t_1, t_2$ and $+$, respectively, with two edges pointing to $+$ from the $t_i$'s. The use of a graph representation rather than of polynomials has two advantages: first, as demonstrated in [20], a graph encoding is more compact as it allows different tuple annotations to share parts of the graph; and second, a graph representation for the operation of the individual modules fits nicely into a graph representation for the provenance of the entire workflow. The resulting graph model that we obtain here is significantly richer than that of [20].

In the remainder of this section we refer to Figure 2(c), which depicts fine-grained provenance for $M_{\text{dealer1}}$ and $M_{\text{agg}}$, and explain in detail how it is generated.

**State nodes.** For each tuple that occurs in the state of some invoked module, we create (1) a p-node labeled with an identifier of this tuple (e.g., node $N_{01}$ for car $C2$ in the example) (2) a p-node of a new type "s" (for "state"), labeled with $\cdot$, to which we connect both the tuple p-node and the module invocation p-node. The $\cdot$ label here has the same meaning of joint dependency as in input / output nodes (see, e.g., node $N_{42}$). State nodes may also be useful in cases where data is shared between modules through the state DB, and not through input-output.

We next formally define provenance propagation for Pig Latin operations. We start with operations that are used in $M_{\text{dealer1}}$, in their order of appearance, showing their effect on the provenance graph. Then, to complete the picture, we define provenance for additional Pig Latin constructs. In what follows we use $v_t$ to refer to the p-node corresponding to the provenance of a tuple $t$.

**`FOREACH` (projection).** [1] Consider the result of `FOREACH A GENERATE f1, f2,...,fn`. For each tuple $t$ in the result, $v_t$ is labeled with a $+$, with incoming edges from each $v_{t'}$ such that $t'$ is in `A`, and its values in attributes `f1, f2,...,fn` are equal to those of $t$.

EXAMPLE 3.1. *The first operation in the query implementing $M_{\text{dealer1}}$ projects over the requested bids given as input to the module (in this case there is only one request, and its provenance is represented by $N_{41}$), to retrieve the requested car model. The tuple obtained as the result of the projection is associated with the $+$-labeled node $N_{50}$.*

**`JOIN`.** For each tuple $t$ in the result of `JOIN A BY f1, B BY f2`, we create a p-node labeled $\cdot$ with incoming edges from $v_{t'}, v_{t''}$, where $t'$ from `A` and $t''$ from `B` join to produce $t$.

EXAMPLE 3.2. *After retrieving the requested car models, $M_{\text{dealer1}}$ continues by joining them with the inventory table. In our case the single requested model is matched to the two cars in the inventory ($C2$ and $C3$). Note that the data on these two cars appears in the inner state of the module, hence state nodes $N_{42}$ and $N_{43}$. The provenance of the two tuples in the result of the join is represented by p-nodes $N_{60}$ and $N_{61}$. Another join is then performed on* `SoldCars`*, but since its result is empty it has no effect on the graph.*

**`GROUP`.** For each tuple $t$ in the result of `GROUP A BY f`, create a p-node labeled $\delta$, with incoming edges from the p-nodes $v_1, ..., v_k$ corresponding to tuples in `A` that have the same grouping attribute value as $t$. [2]

EXAMPLE 3.3. *After finding cars of the requested model, $M_{\text{dealer1}}$ groups them by model. The provenance of the single tuple in the result of this grouping is represented by $N_{71}$. Next, $M_{\text{dealer1}}$ performs* `GROUP` *on the empty* `SoldModelCars` *table, bearing no effect on the graph.*

**`FOREACH` (aggregation).** Recall that `FOREACH` can also be used for aggregation in addition to projection. In this case the provenance of the result is represented as in the case of projection above, but we also represent in the graph the aggregated *value*. To this end we create, for each tuple $t$ in the result, a new v-node labeled with the relevant operation, e.g., `Sum`, `Count`, etc. For each tuple $t'$ in the group corresponding to $t$, we then create a new v-node labeled $\otimes$ and a new v-node labeled with the value $a$ in $t'$ that is being aggregated (if a node for this value does not exist already). We add edges from the $a$-labeled v-node and from $v_{t''}$ to $\otimes$, and from $\otimes$ to the node with the operation name.

EXAMPLE 3.4. *The next step in the module logic is to aggregate the cars of requested models using* `Count`*, computing the number of cars per model. We show a simiplified construction for aggregation, omitting v-nodes that represent tensors and constants. The node representing the single aggregated value in our example is $N_{70}$.*

**`COGROUP`.** For each tuple $t$ in the result of `COGROUP A by f1, B by f2`, create a p-node labeled $\delta$, with incoming edges from p-nodes $v_1, ..., v_k$ (resp. $v_{k+1}, ..., v_l$) corresponding to tuples in `A` (resp. `B`) whose `f1` value (resp. `f2` value) is equal to the grouping attribute value of $t$. As in `GROUP`, tuples in the relations nested in $t$ keep their original provenance.

EXAMPLE 3.5. *$M_{\text{dealer1}}$ next computes* `AllInfoByModel`*, combining request information with the number of available and sold cars. The resulting p-node is $N_{75}$.*

**`FOREACH` (Black Box).** The provenance for a tuple $t$ in the result of `BB(`$t_1$`,...,`$t_n$`)`, where `BB` (Black Box) is a function name, is captured by a node labeled with the function name, to which the $n$ nodes of the inputs $t_1, ..., t_n$ are connected. Depending on the output of the function, the `BB` node may be either a p-node or a v-node.

EXAMPLE 3.6. *$M_{\text{dealer1}}$ next executes* `calcBid`*. The value of the result is represented by the v-node $N_{80}$. Note that this node is connected to $N_{90}$ representing an output tuple, since the computed value is part of this tuple.*

We have described how fine-grained provenance is generated for the operations used in $M_{\text{dealer1}}$. The same operations are used for $M_{\text{agg}}$ so we do not detail the construction.

***Additional Pig Latin operations.*** Fine-grained provenance expressions can similarly be generated for the remaining (non-update) Pig Latin features such as Map datatypes, `FILTER`, `DISTINCT`, `UNION`, and `FLATTEN`, and are omitted due to lack of space. Even joins on attributes with complex types can be modeled by Pig Latin expressions of boolean type. Since relations are unordered in our representation, `ORDER` is not captured in the provenance graph and is a post-processing step. Note that `ORDER` is also a post-processing step in Pig Latin.

# 4. QUERYING PROVENANCE GRAPHS

We next show how fine-grained provenance can be used for supporting complex analysis tasks on workflow executions, *in particular queries that cannot be answered using coarse-grained provenance.*

[1] `FOREACH` can be used for projection (considered now), aggregation and black box invocation (considered later).

[2] This is a "shorthand" for attaching $v_1, ..., v_k$ to a $+$-labeled p-node and then a $\delta$-labeled p-node.

## 4.1 Zoom

Analysts of workflow provenance may be interested in fine-grained provenance for some modules, but in coarse-grained provenance for others. To capture this, we define two transformation operators: ZoomIn and ZoomOut.

ZoomOut of a module hides all of its intermediate computations, as well as its state nodes. We note that, since different invocations of the same module may share state, it does not make sense to ZoomOut from a proper subset of these invocations. For example, if we ZoomOut from $M_{\text{dealer1}}$, then invocations in both the bid request and purchase phases, in all executions of the workflow represented in the provenance graph, must be zoomed-out.

We next show how to identify nodes that represent intermediate computations in invocations of a module $M$.

Definition 4.1. *A node $v$ in a provenance graph $G$ is a part of the intermediate computation of some invocation of a module $M$ iff*
*(1) there exists a directed path $p$ to $v$ from some $v_0 \neq v$, such that $v_0$ is*
*(i) an input node of some invocation of $M$, or*
*(ii) a state node of some invocation of $M$, or*
*(iii) a v-node of some intermediate computation of some invocation of $M$; and*
*(2) there is no output node on $p$ (including $v$).*

Example 4.1. *$N_{60}$ and $N_{70}$ in Figure 2(c) are intermediate computations of an invocation of $M_{\text{dealer1}}$ (the bid phase) because there is a directed path from the input $N_{41}$ to them on which no output node occurs (there is also a directed path to the same nodes from the state p-node $N_{42}$). $N_{101}$ is not an intermediate computation, since all paths to it go through the output node $N_{90}$. The shaded boxes in Figure 2(c) contain the intermediate computations for each module (as well as its input, output, module invocation and state nodes).*

*ZoomOut.* This operation is given two parameters, a provenance graph $G$ and a set of module names $\mathcal{M}$. It returns a new graph $G'$ in which nodes of intermediate computations of modules in $\mathcal{M}$ are removed, and each invocation of $M \in \mathcal{M}$ is represented by a p-node, with the original inputs and output of the module. To ZoomOut on $\mathcal{M}$:

1. Find all the p-nodes of invocations of modules in $\mathcal{M}$.
2. Follow the directed edges from module invocation nodes to find their input and state nodes.
3. According to Definition 4.1, find all the intermediate nodes of invocations of modules in $\mathcal{M}$, remove them and all the edges adjacent to them.
4. Remove the state nodes of invocations of modules in $\mathcal{M}$, and the basic tuple nodes and edges adjacent to those state nodes.
5. For each invocation of $M \in \mathcal{M}$, create a new p-node labeled $M$, connect the invocation inputs to it and connect it to the invocation outputs.

Applying ZoomOut on all modules in a fine-grained provenance graph $G$ results in a coarse-grained provenance graph.

*ZoomIn.* ZoomIn is the inverse of ZoomOut, namely $ZoomIn(ZoomOut(G, M), M) = G$.

Example 4.2. *Consider the provenance graphs in Section 3 (coarse-grained in Figure 2(b) and fine-grained in Figure 2(c)). Observe that the latter is obtained from the former by zooming into the modules $M_{\text{dealer1}}$ and $M_{\text{agg}}$.*

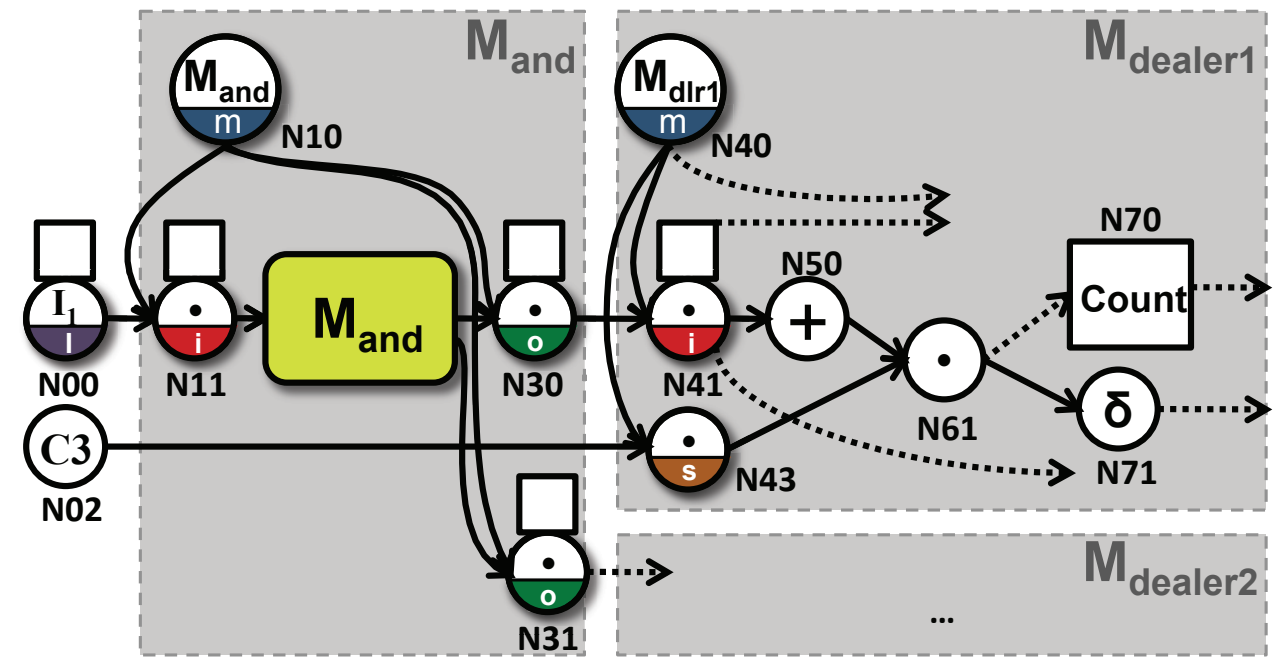

**Figure 3: Propagating the deletion of $C2$.**

To conclude, we note that a different semantics of zoom operations was introduced in the context of coarse-grained provenance in [4], where the provenance of multiple modules is abstracted away using a composite module. Our notion of zoom is different and more complex due to the maintenance of fine-grained provenance, and in particular of module state that may be shared across multiple executions.

## 4.2 Deletion Propagation

Another application of fine-grained provenance is to analyze how potential deletions propagate through the workflow execution, allowing users to assess the effect that tuple $t$ has on the generation of some other tuple $t'$. Intuitively, deletion of a tuple $t$ propagates to all tuples whose existence depends on $t$, i.e., all tuples whose provenance has a multiplicative factor (or a single additive factor) dependent on the annotation of $t$. The process continues recursively, since additional tuples may now have no derivations. More formally,

Definition 4.2. *The result of deleting a node $v$ from a provenance graph $G$ is a graph $G'$ obtained from $G$ by removing $v$ and all edges adjacent to it, and then repeatedly removing every node (and all edges adjacent to it) that* either *(1) all of its incoming edges were deleted* or *(2) is labeled with $\cdot$ or $\otimes$ and one of its incoming edges was deleted.*

We note that the result of a deletion may not correspond to the provenance of any actual workflow execution, but it may be of interest for analysis purposes.

Example 4.3. *Consider first a query that analyzes the effect of removing car $C2$ from stock. Propagating its deletion, we obtain the graph in Figure 3. Note that the* COUNT *aggregate is now applied to a single value (the one obtained for car $C3$), and so we can easily re-compute its value.*

Example 4.4. *Consider the deletion of user request $I_1$ (node $N_{00}$), and observe that its propagation results in the deletion of the entire graph, except for nodes standing for state tuples or module invocations. Intuitively, if no bid request were submitted the execution would not have occurred.*

## 4.3 Other Queries Enabled

Since provenance is represented as a graph that captures fine-grained, database-style operations on input and state, along with coarse-grained module invocations, users can ZoomIn/ZoomOut to a chosen level of detail and then issue queries in the graph language of their choice (e.g. ProQL [20]) augmented with deletion propagation. In particular, *dependency queries* are enabled, i.e. queries that ask, for a pair of nodes $n, n'$, if the existence of $n$ depends on that of $n'$. This may be answered by checking for the existence of $n$ in the graph obtained by propagating the deletion of $n'$. This can be further extended to *sets* of nodes.

EXAMPLE 4.5. *Continuing with our running example, we observe that the calculation of the bid does not depend on the existence of car $C2$, since the bid tuple still exists in the graph obtained by propagating the deletion of node $N_{01}$ corresponding to $C2$. In contrast, in Example 4.4, bid calculation does depend on the existence of tuple $I_1$ (node $N_{00}$).*

Examples of other analytic queries that are now enabled were given in the Introduction.

# 5. EXPERIMENTS

We now describe *Lipstick*, a prototype that implements provenance tracking and supports provenance queries. We present the architecture of *Lipstick* in Section 5.1, and describe *WorkflowGen*, a benchmark used to evaluate the performance of *Lipstick*, in Section 5.2. Section 5.3 outlines our experimental methodology. We show that tracking provenance during workflow execution has manageable overhead in Section 5.4, and that provenance graphs can be constructed and queried efficiently in Sections 5.5 and 5.6.

## 5.1 System Architecture

*Lipstick* consists of two sub-systems: *Provenance Tracker* and *Query Processor*, which we describe in turn.

**Provenance Tracker.** This sub-system is responsible for tracking provenance for tuples that are generated over the course of workflow execution, based on the model proposed in this paper. The sub-system output is written to the file-system, and is used as input by the *Query Processor* sub-system, described below. We note that *Provenance Tracker* does not involve any modifications to the Pig Latin engine. Instead, it is implemented using Pig Latin statements ( some of which invoke user-defined functions implemented in Java) that are invoked during workflow execution.

**Query Processor.** This sub-system is implemented in Java and runs in memory. It starts by reading provenance-annotated tuples from disk and building the provenance graph. In our current implementation, we store information about parents and children of each node, and compute ancestor and descendant information as appropriate at query time. An alternative is to pre-compute the transitive closure of each node, or to keep pair-wise reachability information. Both these options would result in higher memory overhead, but may speed up query processing.

Once the graph is memory-resident, we can execute queries against it. Our implementation supports zoom (Section 4.1), deletion (Section 4.2) and subgraph queries. A subgraph query takes a node id as input and returns a subgraph that includes all ancestors and descendants of the node, along with all siblings of its descendants. The result of this query may be used to implement dependency queries (Section 4.3).

## 5.2 Evaluation Benchmark

We developed a benchmark, called *WorkflowGen*, that allows us to systematically evaluate the performance of *Lipstick* on different types of workflows. *WorkflowGen* generates and executes two kinds of workflows, described next.

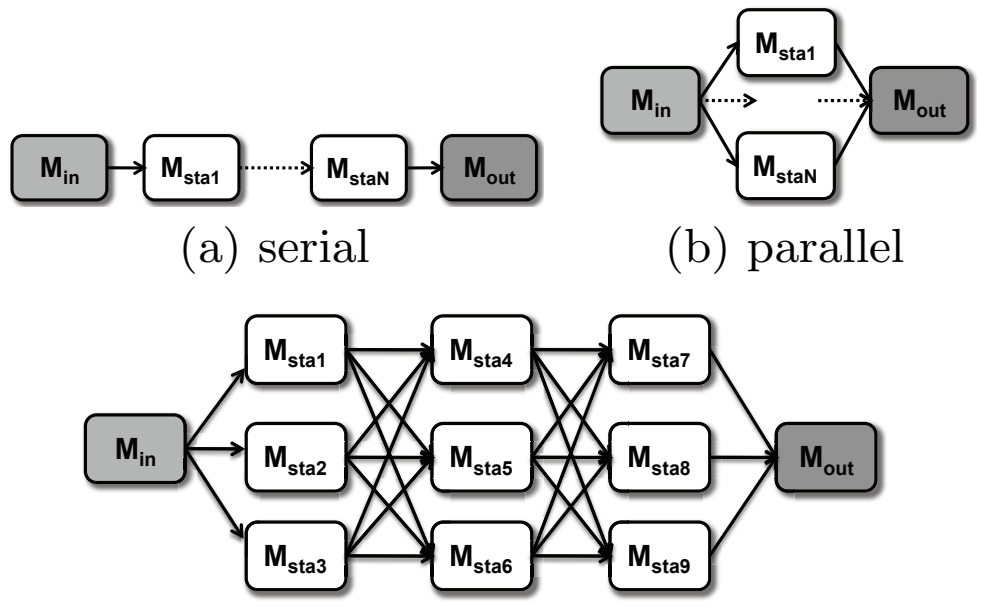

(a) serial (b) parallel

(c) dense, fan-out 3, 9 station modules

**Figure 4: Sample *Arctic stations* workflows.**

**Car dealerships.** This workflow, which was used as our running example, has a fixed topology, with four car dealership modules executing in parallel. Already this simple workflow demonstrates interesting features such as aggregation, black box invocation and intricate tuple dependencies.

*WorkflowGen* executes the *Car dealerships* workflow as follows. A single *run* of a workflow is a series of multiple consecutive executions and corresponds to an instance of the provenance graph. Each dealership starts with the specified number of cars (*numCars*), with each car randomly assigned one of 12 German car models. A buyer is fixed per run; it is randomly assigned a desired car model, a reserve price and a probability of accepting a bid. A run terminates either when a buyer chooses to purchase a car, or the maximum number of executions (*numExec*) is reached.

**Arctic stations.** *WorkflowGen* also implements a variety of workflows that model the operation of meteorological stations in the Russian Arctic, and is based on a real dataset of monthly meteorological observations from 1961-2000 [27]. Workflows in this family vary w.r.t. the number of *station* modules, which ranges between 2 and 24. In addition to station modules, each workflow contains exactly one input and one output module. Workflows also vary w.r.t. *topology*, which is one of *parallel*, *serial*, or *dense*. Figure 4 presents specifications for several workflow topologies. For *dense* workflows we vary both the number of modules and the fan-out; Figure 4(c) shows a representative workflow.

The input module $M_{in}$ receives three inputs: current *year* and *month*, and query *selectivity* (one of *all*, *season*, *month*, or *year*); these are passed to each station module $M_{stai}$. Each $M_{stai}$ has its database state initialized with actual historical observations for one particular Arctic station from [27]. In addition to workflow inputs, $M_{stai}$ also receives a value for minimum air temperature (*minTemp*) from each module $M_{staj}$ from which there is an incoming edge to $M_{stai}$. So, $M_{sta5}$ in Figure 4(c) gets three *minTemp* values as input, one from each $M_{sta1}$, $M_{sta2}$ and $M_{sta3}$. $M_{stai}$ starts by taking a measurement of six meteorological variables, including air temperature, and recording it in its internal state. Next, $M_{stai}$ computes the lowest air temperature that it has observed to date (as reflected in its state) for the given selectivity. For example, if selectivity is *all*, the minimum is taken w.r.t. all historical measurements at the station, if it is *season*, then measurements for the current season (over all years) are considered (selectivity of $\frac{1}{4}$), etc. Finally, $M_{stai}$ computes the minimum of its locally computed lowest temperature and of *minTemp* values received as input, and outputs the result. The output module $M_{out}$ computes and outputs the over-all minimum air temperature.

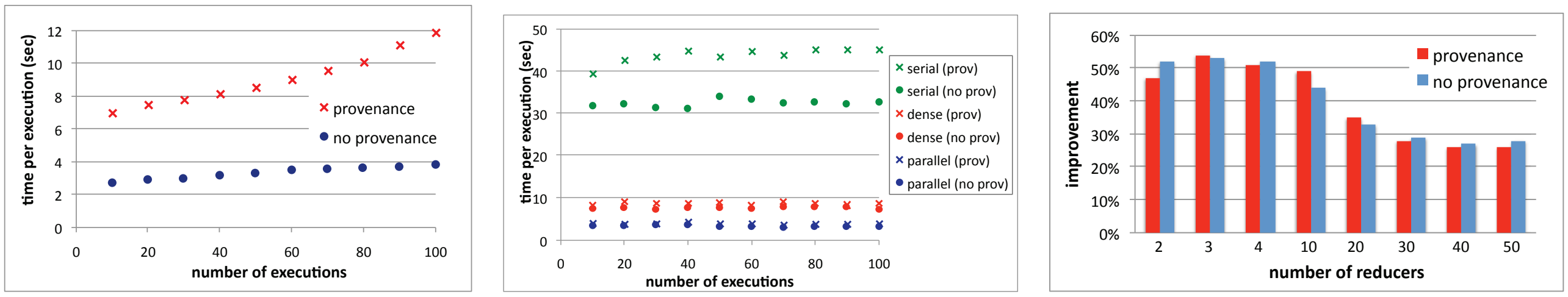

(a) *Car dealerships*, local mode (b) *Arctic stations*, local mode (c) *Car dealerships*, impact of parallelism

**Figure 5: Pig Latin workflow execution time.**

*Arctic stations* workflows allow us to measure the effect of workflow size and topology on the cost of tracking and querying provenance. Selectivity, supplied as input, has an effect on the size of the provenance of the intermediate and output tuples computed by each workflow module.

## 5.3 Experimental Methodology

Experiments in which we evaluate the performance of *Provenance Tracker* are implemented in Pig Latin 0.6.0. Hadoop experiments were run on a 27-node cluster running Hadoop 0.20.0. All other experiments were executed on a MacBook Pro running Mac OS X 10.6.7, with 4 GB of RAM and a 2.66 GHz Intel Core i7 processor.

All results are averages of 5 runs per parameter setting, i.e., 5 execution histories are generated for each combination of $numCars$ and $numExec$ for *Car dealerships*, and for each topology, number of modules, selectivity, and $numExec$ for *Arctic stations*. For each run we execute each operation 5 times, to control for the variation in processing time.

## 5.4 Tracking Provenance

We now evaluate the run-time overhead of tracking provenance, which occurs during the execution of a Pig Latin workflow in *Lipstick*. We first show that collecting provenance in local mode is feasible, and then demonstrate that provenance tracking can take advantage of parallelism.

Figure 5(a) presents the execution time of *Car dealerships* with 20,000 cars (5000 cars per dealership), in *local mode*, as a function of the number of prior executions of the same workflow (i.e., $numExec$ per run). We plot performance of two workflow versions: with provenance tracking and without. The number of prior executions increases the size of state over which each dealership in the workflow reasons while generating a bid. Therefore, as expected, execution time of the workflow increases with increasing number of prior executions. Tracking provenance does introduce overhead, and the overhead increases with increasing number of historical executions. For example, in a run in which the dealership is executed 10 times (10 bids per dealership), 2.7 sec are needed per execution on average when no provenance is recorded, compared to 7 sec with provenance. With 100 bids per dealership, 3.8 sec are needed on average without provenance, compared to 11.9 sec with provenance.

Figure 5(b) show results of the same experiment for three *Arctic stations* workflows, with parallel, serial, and dense topologies, all with 24 station modules. The dense workflow has fan-out 6, executing 6 station modules in parallel. Module selectivity was set to *month* in all cases, i.e., the minimum air temperature was computed over $\frac{1}{12}$ of the state tuples. Observe that parallel workflow executes fastest, followed by dense, and then by serial. This is due to the particulars of our implementation, in which all modules running in parallel are implemented by a single Pig Latin program, while each module in the serial topology (and each set of 6 modules in the dense topology) are implemented by separate Pig Latin programs, with parameters passed through the file system. (This is true of our implementation of *Arctic stations* workflows both with and without provenance tracking.) Observe also that tracking provenance introduces an overhead of 16.5% for parallel, 20.0% for dense, and 35% for serial topologies. Finally, note that there is no increase in execution time of the workflows, either with or without provenance tracking, with increasing $numExec$. This is because there is no direct dependency between current and historical workflow outputs. The provenance of intermediate and output tuples does increase in size, because new observations are added to the state, but this does not have a measurable effect on execution time.

In the next experiment, we show that workflows that track provenance can take full advantage of parallelism provided by Hadoop. We control the degree of parallelism (the number of reducers per query) by adding the $PARALLEL$ clause to Pig Latin statements. We execute this experiment on a 27-node Hadoop cluster with 2 reducer processes running per machine, for a total of up to 54 reducers. Results of our evaluation for *Car dealerships* are presented in Figure 5(c) and show the percent improvement of executing the workflow with additional parallelism in the reduce phase, compared to executing it with a single reducer.

Best improvement is achieved with between 2 and 4 reducers, and is about 50% both with and without provenance. This is because the part of our workflow that lends itself well to parallelization is when 4 bids are generated, one per dealership. However, there is a trade-off between the gain due to parallelism (highest with 4 reducers) and the overhead due to parallelism (also higher with 4 reducers than with 2 and 3). 3 reducers appear to hit the sweet spot in the trade-off, although performance with between 2 and 4 reducers is comparable. Note that, although we are able to observe clear trends, small differences, e.g., % improvement with provenance tracking vs. without, for the same number of reducers, are due to noise and insignificant.

**In summary,** tracking provenance as part of workflow execution does introduce overhead. The amount of overhead, and whether or not overhead increases with time, depends on workflow topology and on the functionality of workflow modules, e.g., the extent to which they modify internal state, use aggregation and black-box functions. Nonetheless, the overhead of tracking provenance is manageable for the workflows in our benchmark. Furthermore, since *Lipstick* is implemented in Pig Latin, it can take full advantage of Hadoop parallelism, making it practical on a larger scale.

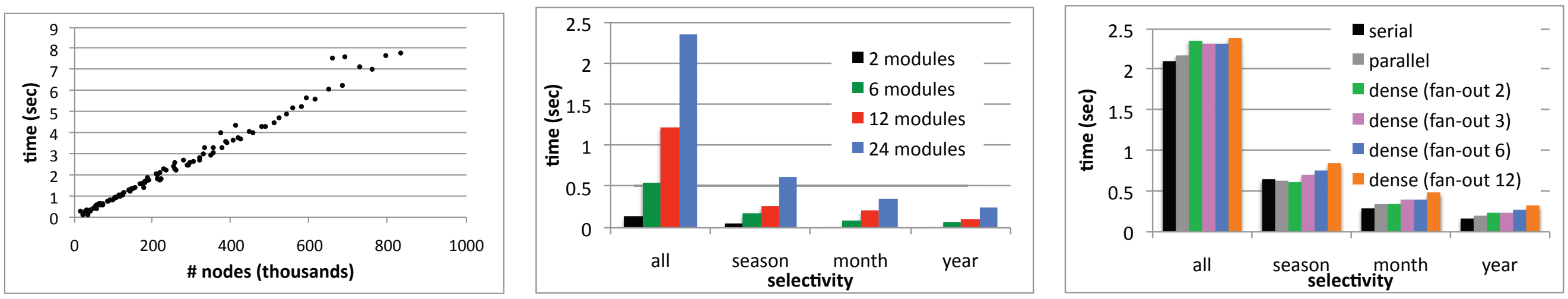

(a) *Car dealerships* (b) *Arctic stations* dense, fan-out 2 (c) *Arctic stations*, 24 modules

**Figure 6: Provenance graph building time.**

## 5.5 Building the Provenance Graph

In this section we consider the time it takes to build the provenance graph in memory from provenance-annotated tuples. Figure 6 summarizes our results.

First we note that, for a fixed size of initial state, the number of provenance graph nodes increases approximately linearly with the number of workflow executions. (Plots omitted due to lack of space.) Figure 6(a) plots graph building time as a function of the number of graph nodes for *Car dealerships*. Observe that time is linear in the number of nodes, and is below 8 sec in all cases. Similar results were observed for *Arctic stations*.

Figures 6(b) and 6(c) show the effect of workflow size and topology on provenance graph building time. All workflows were executed 100 times to a run. Figure 6(b) gives results for *Arctic stations* workflows with dense topology and with fan-out 2, for different number of modules. Observe that, as expected, graph building time increases with increasing number of modules. A station module uses the selectivity to determine w.r.t. which observations to compute the minimum temperature (an intermediate tuple that contributes to module output). Selectivity *all* is lowest (all state tuples are considered), *season* corresponds to $\frac{1}{4}$ of the tuples, *month* — to $\frac{1}{12}$ of the tuples, and *year* — to at most 12 tuples. The lower the selectivity — the more edges there are in the provenance graph, and the more expensive it is to construct, as reflected in the plots in Figures 6(b) and 6(c). Based on Figure 6(c) we observe that workflow building time does not vary significantly across topologies, but appears to be shortest for serial workflows, followed by parallel, and then by dense, in increasing order of fan-out.

Finally, let us consider the size of the provenance of the output tuples and demonstrate that provenance in our case is indeed fine-grained, i.e., that workflow outputs depend only on a portion of the input and state tuples. In the *Car dealerships* workflow, output tuples correspond to sold cars; input tuples are bid requests and user choices; internal state is the total number of cars in the four dealerships. For $numCars$ = 20,000 and $numExec$ = 10,000, we observed that any particular output tuple depends on between 1.8% and 2.2% of the state tuples (415 tuples on average) and on two input tuples. In contrast, if one were to use traditional coarse-grained provenance [23], each sale would depend on 100% of the state tuples (20,000 tuples in this example) and on all user inputs (10,000 in our experiment).

**In summary**, we conclude that building a provenance graph is feasible for workflows of various sizes and topologies. This may be done as part of an off-line process, e.g., during initialization of a query or visualization interface.

## 5.6 Processing Provenance Queries

In this section we show that the provenance queries described in Section 4 can be evaluated efficiently.

**Zoom.** Figure 7(a) shows the performance of ZoomOut (Section 4.1) as a function of provenance graph size. Here, $numCars$ = 20,000; $numExec \in$ [1000, 10,000]. ZoomOut replaces each occurrence of an execution of a module with a meta-node, hiding internal nodes. We show results for `dealer` and `aggregate`. We observe that execution time is linear in graph size for both modules, and that zooming out on `aggregate` is faster (at most 0.68 sec) followed by dealer (at most 2.02 sec). The main difference between `dealer` and `aggregate` is the number of *module instances*: `dealer` executes at most 5 times per workflow execution (once per dealership in the bidding phase, and at most once in the purchase phase), while `aggregate` executes at most once. In this experiment, `dealer` had between 3000 and 115,000 instances, while `aggregate` had between 750 and 38,000 instances. For ZoomIn we observe similar trends as for ZoomOut: run-time is linear in graph size, and is fastest for `aggregate` followed by `dealer`. ZoomIn is about three times faster than ZoomOut, taking at most 0.68 sec for `dealer` and 0.23 for `aggregate`.

**Subgraph.** A subgraph query takes node id as input and returns a subgraph that includes the node's ancestors, descendants, and siblings of descendants. To observe performance trends, we select nodes that we expect to induce large subgraphs, choosing 50 nodes with the highest number of children per run. Figure 7(b) plots query processing time as a function of the number of nodes in the resulting subgraph for *Car dealerships*, with $numCars$ =20,000 and $numExec \in$ [1000, 10,000]. Among 50 nodes chosen per run, only between 5 and 12 nodes had measurable (over 1 ms) query execution time. Observe that the subgraph is computed in under 0.2 sec in all cases, with the highest processing time corresponding to a subgraph with over 40 thousand nodes, and that processing time increases approximately linearly in subgraph size.

Figure 7(c) plots query execution time for subgraph for *Arctic stations* with 24 modules. This plot shows that query time depends on selectivity, which influences the numbers of nodes and edges in the provenance graph. Query time also depends on module topology, because it results in a different number of graph edges. For a fixed selectivity, all graphs have the same number of nodes but differ on (1) the number of in-coming edges for module and workflow output nodes (higher for higher fan-out) and (2) the number of high-degree nodes on a path to workflow output (lower for higher fan-out). Due to this trade-off, dense workflows with fan-out 3 have highest query execution time.

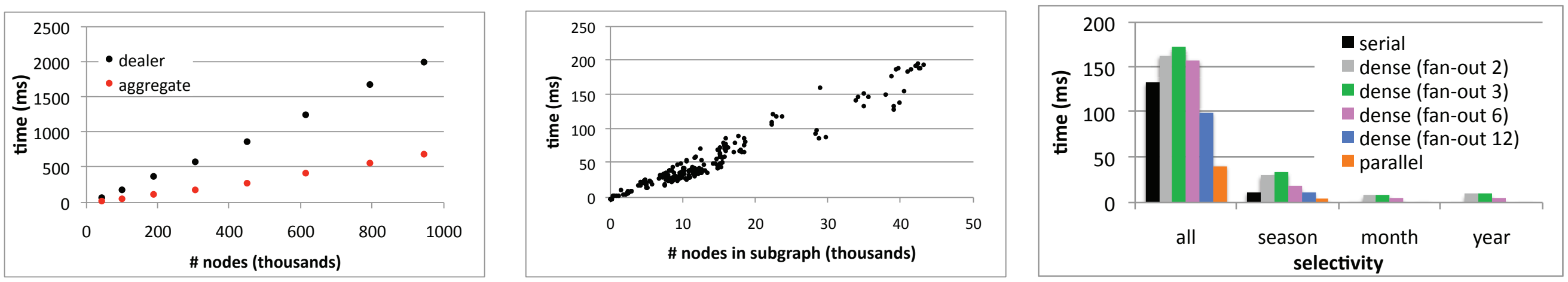

(a) ZoomOut, *Car dealerships* (b) Subgraph, *Car dealerships* (c) Subgraph, *Arctic stations*, 24 modules

**Figure 7: Query processing time.**

**Delete.** This query takes node id as input, and propagates deletion of the node to its descendants. Because there is no need to look at ancestors of a node, this query traverses a much smaller subgraph than a subgraph query. We selected 50 nodes per run in the same way as for subgraph, and observed that processing times were lower than 1 ms in most cases, and at most 10-13 ms per node. Because delete queries are so efficient, we were unable to observe any trends that relate graph or subgraph size to processing time.

**In summary,** we showed that the provenance graph can be queried efficiently, with sub-second execution times in the vast majority of tested cases.

# 6. CONCLUSIONS

In this paper we studied fine-grained provenance for workflows. We focused on workflows where the individual modules are implemented in Pig Latin, gave a provenance model for Pig Latin queries, and showed how the model may be extended to capture entire workflow executions. We further demonstrated the usefulness of the approach by showing how it allows to answer practical questions that are of interest to analysts of workflows, and that cannot be answered using standard coarse-grained workflow provenance.

Our focus was on developing a model for fine-grained provenance, and on showing its feasibility by means of an experimental evaluation. As future work we will consider optimization techniques for model representation and for query algorithms. We will also study extensions of the queries suggested here, towards a full-fledged query language. Finally, we will extend our model to account for additional common features of workflows, such as fork and loop constructs.

# 7. ACKNOWLEDGMENTS

This work was supported in part by the US National Science Foundation grants IIS-0803524 and IIS-0629846, and grant 0937060 to the Computing Research Association for the CIFellows Project, by the Israel Science Foundation, and by the US-Israel Binational Science Foundation.